\documentclass[aps,nofootinbib,preprint]{revtex4}
\usepackage{graphicx,color}
\usepackage{epsfig}
\usepackage{amsmath}
\usepackage{amssymb}

\begin{document}

\title{Lifetime of doubly charmed baryons}
\author{Chao-Hsi Chang$^{1,2}$ \footnote{email:
zhangzx@itp.ac.cn}, Tong Li$^{3}$\footnote{email:
allongde@mail.nankai.edu.cn}, Xue-Qian Li$^{3}$\footnote{email:
lixq@nankai.edu.cn} and Yu-Ming Wang$^{4}$\footnote{email:
wangym@mail.ihep.ac.cn}}
\address{$^1$CCAST (World Laboratory), P.O.Box 8730, Beijing 100080,
P.R. China\\
$^2$Institute of Theoretical Physics, Chinese Academy of Sciences,
P.O.Box 2735, Beijing 100080, P.R. China\\
$^3$ Department of Physics, Nankai University, Tianjin, 300071,
P.R. China\\
$^4$Institute of High Energy Physics, Chinese Academy of Sciences,
P.O.Box 918, Beijing 100049, P.R. China}

\begin{abstract}
In this work, we evaluate the lifetimes of the doubly charmed
baryons $\Xi_{cc}^{+}$, $\Xi_{cc}^{++}$ and $\Omega_{cc}^{+}$. We
carefully calculate the non-spectator contributions at the quark
level where the Cabibbo-suppressed diagrams are also included. The
hadronic matrix elements are evaluated in the simple
non-relativistic harmonic oscillator model. Our numerical results
are generally consistent with that obtained by other authors who
used the diquark model. However, all the theoretical predictions
on the lifetimes are one order larger than the upper limit set by
the recent SELEX measurement. This discrepancy would be clarified
by the future experiment, if more accurate experiment still
confirms the value of the SELEX collaboration, there must be some
unknown mechanism to be explored.

\end{abstract}
\maketitle

\section{Introduction}

The quite large difference of the lifetimes between $D^{\pm}$ and
$D^0$ and the lifetimes close to each other for $B^{\pm}$ and
$B^0$ are well explained by taking into account the non-spectator
effects\cite{Bigi1}. This success implies that the mechanism which
governs the reactions at quark level is well understood. When we
apply the mechanism to the heavy baryon case, some problems
emerge. The famous puzzle in the heavy-flavor field that the
lifetime of $\Lambda_b$ is remarkably shorter than that of $B$
meson is much alleviated recently when the operators of higher
dimensions are taken into account\cite{Franco,Gabbiani}. The more
recent experimental value of the ratio $\tau(\Lambda_b)/\tau(B^0)
=1.041\pm 0.057$\cite{Lambdabexp} is close to the theoretical
evaluation\cite{Gabbiani}. However, in the theoretical works, one
can notice that the evaluation of hadronic matrix elements is
still very rough and based on some approximations. The possible
errors brought up by the uncertainties in the hadronic matrix
elements are still uncontrollable. In our recent work\cite{He1},
we find that the short-distance contributions to the branching
ratio of $\Lambda_b\rightarrow \Lambda \gamma$ which is evaluated
in the PQCD approach, are much smaller than that from
long-distance effects. Therefore, even though one has a full
reason to believe that the low-energy QCD should solve the
discrepancy if it exists, he must find a proper way to deal with
the hadronic matrix elements.

The observation of doubly charmed baryon $\Xi_{cc}^{+}$ by the
SELEX Collaboration at FERMILAB\cite{SELEX} provides an
opportunity to investigate the hidden problems. Hopefully the
study may shed some lights on the unknown non-perturbative QCD
effects which result in obvious difference between baryons and
mesons. Because $\Xi_{cc}^{+}$ contains two heavy quarks, by the
heavy quark effective theory (HQET) the situation may become
relatively simple and clear compared to the case of $\Lambda_b$ or
$\Lambda_c$ which possesses only one heavy quark. Thus a careful
study on the $\Xi_{cc}^{+}$ is necessary and interesting. Several
groups already investigated the two-heavy-flavor baryons a long
time ago\cite{Kiselev,Guberina}. In their work, the evaluation of
the hadronic matrix elements is based on the quark-diquark
structure of the baryons. This is definitely reasonable, it is
believed that two heavy quarks can constitute a more stable and
compact color-anti-triplet diquark\cite{Wise}. However, since
charm quark, even b-quark, is not so heavy that the degree of
freedom of the light flavor can be ignored, the diquark scenario
may bring up certain errors, especially when evaluating lifetimes
of baryons, because only inclusive processes are concerned. In
this work, we do not use the diquark picture, but instead, adopt a
simpler non-relativistic model for the baryon and re-evaluate the
hadronic matrix elements. As a by-product, one can compare the
results by the diquark picture with that by the three
valence-quark picture. It may help us to better understand the
diquark picture and its application range.

The advantage is obvious, that we only concern the inclusive
processes in terms of the optical theorem when calculating the
lifetime. Therefore, we do not need to deal with the hadronization
to light hadrons. The only non-perturbative effects come from the
wave function of the heavy baryon. Moreover, since there are two
heavy quarks in the baryon, the relativistic effects are not so
significant and the framework of non-relativistic harmonic
oscillator model might lead to a reasonable result.

Moreover, at the quark level, we carry out similar calculations as
that in the literature, but we keep some new operators which are
CKM suppressed and contribute to the lifetime. They appear at the
non-spectator scattering at order of $\frac{1}{m_c^3}$ in heavy
quark expansion(HQE). Later, our numerical results show that their
contributions are indeed very tiny to make any substantial
contributions.

All the concerned parameters in the model are obtained by fitting
data, therefore we avoid some theoretical uncertainties and obtain
reasonable results. Comparing these results with data, we may gain
information about the the whole picture.

Our paper is organized as follows. In Section.II we derive the
formulation for the lifetimes of $\Xi_{cc}^{+}$, $\Xi_{cc}^{++}$
and $\Omega_{cc}^+$ which include the non-spectator effects. In
Section.III, we use a simple model, i.e. the harmonic oscillator,
to estimate the hadronic matrix elements. In Section.IV we present
our numerical results along with the values of all the input
parameters. The last section is devoted to our conclusion and
discussion.

\section{Formulation for Lifetimes of $\Xi_{cc}^{+}$,
$\Xi_{cc}^{++}$, $\Omega_{cc}^{+}$ }
\subsection{Spectator Contribution to  Lifetimes of  $\Xi_{cc}^{+}$,
$\Xi_{cc}^{++}$, $\Omega_{cc}^{+}$ }

The lifetime is determined by the inclusive decays. Thus one can
use the optical theorem to obtain the total width (lifetime) of
the heavy hadron by calculating the absorptive part of the
forward-scattering amplitude.

The total width is then written as
\begin{eqnarray}
\mathbf{\Gamma}(H_{Q}\rightarrow X)=\frac
{1}{m_{H_{Q}}}\text{Im}\int d^{4}x\langle
H_{Q}|\hat{T}|H_{Q}\rangle =\frac {1}{2 m_{H_{Q}}}\langle
H_{Q}|\hat{\Gamma}|H_{Q}\rangle,
\end{eqnarray}
where
\begin{eqnarray}
\hat{T}=T\{i\mathcal{L}_{eff}(x),\mathcal{L}_{eff}(0)\}
\end{eqnarray}
and $\mathcal{L}_{eff}$ is the relevant effective Lagrangian.
$1/m_{Q}$ is the expansion parameter, and the non-local operator
$\hat{T}$ is expanded as a sum of local operators and the
corresponding Wilson coefficients include terms with increasing
powers of $1/m_{Q}$. Definitely, the lowest dimensional term
dominates in the limit $m_{Q}\rightarrow \infty$ and it is the
dimension-three operator $\bar{c}c$. The total width of a charmed
hadron $H_{c}$ is determined by $\text{Im}\langle H_c|
\hat{T}|H_c\rangle$\cite{B Decays} with a proper
normalization\cite{Wuyl}.
 \begin{eqnarray}
 \Gamma(H_{c}\rightarrow f)&=&\frac{G_{F}^{2}m_{c}^{5}}
 {192\pi^{3}}|V_{CKM}|^{2}\{ c_{3}(f)\langle H_{c}|
 \bar{c}c |H_{c}\rangle\nonumber \\
 &&+c_{5}(f)\frac{\langle H_{c}|
 \bar{c} i \sigma_{\mu \nu} G^{\mu \nu} c |H_{c}\rangle}{m_{c}^{2}}+\nonumber \\
 && \sum_{i}c_{6}^{(i)}(f)
 \frac{ \langle H_{c}|
 (\bar{c}\Gamma_{i} q) (\bar{q} \Gamma_{i} c) |H_{c}\rangle}{m_{c}^{3}}
 +\mathcal{O}(\frac{1}{m_{c}^{4}})\},
 \end{eqnarray}
where the coefficients $c_{i}(f)$ depend on the masses of the
internal quarks in the loop. The coefficient $c_{3}(f)$ has been
calculated to one-loop order\cite{Hokim,Nir,Bagan} whereas the
coefficient $c_{5}(f)$ is evaluated at the tree
level\cite{Bigi2,Falk}. $V_{CKM}$ is the Cabibbo-Kabayashi-Maskawa
mixing matrix elements and $G_{\mu\upsilon}$ is the gluonic field
strength tensor. Since the third term involves light quarks, it
can be different for charmed hadrons with various light flavors.
Thus, the difference appears at the $1/m_{c}^3$ order and in the
hadronic matrix elements of four-quark operators. The
contributions at orders higher than $1/m_{c}^3$ are neglected.

To the lowest order, the main contribution comes from the heavy
quark(charm quark) decays, while the light flavors are treated as
spectators. The contributions are due to the semileptonic and the
nonleptonic decays as follows:
\begin{eqnarray}
&&\Gamma(c\rightarrow s)=\sum_{l=e,\mu}\Gamma_{c\rightarrow s
\bar{l}\upsilon}+\sum_{q(q')=u,d,s} \Gamma_{c\rightarrow s
\bar{q}q'}
\end{eqnarray}
The semileptonic and nonleptonic decay rates of the $c$ quark up
to order $1/m_{c}^{2}$ has been evaluated by many
authors\cite{H.Y.Cheng1}, and here we would directly use their
results.

\subsection{Non-spectator Contributions to Inclusive Decays of $\Xi_{cc}^{+}$,
$\Xi_{cc}^{++}$, $\Omega_{cc}^{+}$} The total width of hadrons which
involve at least one charm quark $c$ can be decomposed into two
parts£º
\begin{eqnarray}
&&\Gamma(H_{Q} \rightarrow
f)=\Gamma^{spectator}+\Gamma^{nonspectator}.
\end{eqnarray}
For the spectator scenario, the contribution to the total width of
the ($ccd$)-baryon ground state $\Xi_{cc}^{+}$, the ($ccu$)-baryon
ground state $\Xi_{cc}^{++}$ and the ($ccs$)-baryon ground state
$\Omega_{cc}^{+}$ should be a sum of decays rates of two $c-$quarks
individually£¬namely
\begin{eqnarray}
&&\Gamma_{ccq}^{spec}\simeq 2\Gamma_{c}^{spec}\,,\;\;\;\; q=u,d,s.
\end{eqnarray}
To derive the non-spectator contributions for decays of
$\Xi_{cc}^{+}$, $\Xi_{cc}^{++}$ and $\Omega_{cc}^{+}$, we need the
relevant effective Lagrangian:\cite{Buccella}
\begin{eqnarray}
\mathcal{L}_{eff}^{(\Delta
c=1)}(\mu=m_{c})&=&-\frac{4G_{F}}{\sqrt{2}} \{V_{cs}
V_{ud}^{*}[C_{1}(\mu)\bar{s}\gamma^{\mu}Lc \bar{u}{\gamma_\mu}Ld+
C_{2}(\mu)\bar{u}\gamma^{\mu}Lc
\bar{s}{\gamma_\mu}Ld ]\nonumber \\
&&+V_{cd} V_{ud}^{*}[C_{1}(\mu)\bar{d}\gamma^{\mu}Lc
\bar{u}{\gamma_\mu}Ld+  C_{2}(\mu)\bar{u}\gamma^{\mu}Lc
\bar{d}{\gamma_\mu}Ld ]\nonumber \\
&&+V_{cs} V_{us}^{*}[C_{1}(\mu)\bar{s}\gamma^{\mu}Lc
\bar{u}{\gamma_\mu}Ls+  C_{2}(\mu)\bar{u}\gamma^{\mu}Lc
\bar{s}{\gamma_\mu}Ls ]\nonumber \\
&&+V_{cs}\sum_{l=e,\mu}\bar{s}\gamma_{\mu}Lc
\bar{\nu_{l}}\gamma^{\mu} L l\}+h.c.
\end{eqnarray}
where $L$ denotes ${1-\gamma_5\over 2}$.\\
(i) The inclusive decays of $\Xi_{cc}^{+}$:\\
There are four diagrams which contribute to the the width of
$\Xi_{cc}^{+}$, as shown in Fig.1. Here we also include the
Feynman diagrams which are CKM suppressed. Fig 1.(a),(c) are the
W-exchange diagrams (WE), while Fig 1.(b),(d) are the
pauli-interference diagrams (PI). Here Fig 1.(d) is arisen from
the semi-leptonic decay of the charm quarks with the $d-$quark in
$\Xi_{cc}^{+}$. For the WE-type diagrams, we derive the
contribution to the width as
\begin{eqnarray}
\hat{\Gamma}_{WE}^{\Xi_{cc}^{+}}&=&\frac{2G_{F}^{2}}{\pi}
(|V_{cs}|^{2} |V_{ud}|^{2}C(z_{s+},z_{u+}) +|V_{cd}|^{2}
|V_{ud}|^{2}C(z_{u+},z_{d+}))P_{+}^{2}\nonumber \\
&&\{[C_{1}^{2}(\mu)+C_{2}^{2}(\mu)] \bar{c}\gamma^{\mu}Lc
\bar{d}{\gamma_\mu}Ld+ 2 C_{1}(\mu)C_{2}(\mu)
\bar{c}\gamma^{\mu}Ld \bar{d}{\gamma_\mu}Lc\},
\end{eqnarray}
where $P_{+}=p_{c}+p_{d}$,
$z_{q+}=\frac{m_{q}^{2}}{P_{+}^{2}}(q=u,d,s)$. The definition of
the function $C(z_1,z_2)$ is
\begin{eqnarray}
C(z_1,z_2)=-[-2(x_2^3-x_1^3)-(x_2^2-x_1^2)(3+2z_1-2z_2)+4z_1(x_2-x_1)],
\end{eqnarray}
where $x_{1,2}=\frac{(1+z_1-z_2)\mp\sqrt{(1+z_1-z_2)^2-4z_1}}{2}$.
In the expressions $q $ and $\bar{q}$ are free field opearotors of
quark and antiquark, and we will show in next section that all the
non-perturbative QCD effects are included in the wavefunctions.
Their explicit expressions are given as
\begin{eqnarray}
&&q=\int \frac {d^{3}k}{(2\pi)^{3}} \frac
{m_{q}}{E_{q}}\sum_{\alpha=1,2}\left(b_{q_{\alpha}}(k)
u_{q}^{\alpha}(k)e^{-ikx}+d_{q_{\alpha}}^{+}(k)
\upsilon_{q}^{\alpha}(k)e^{+ikx}\right)\\
&&\bar{q}=\int \frac {d^{3}k}{(2\pi)^{3}} \frac
{m_{q}}{E_{q}}\sum_{\alpha=1,2}\left(b_{q_{\alpha}}^{+}(k)
\bar{u}_{q}^{\alpha}(k)e^{ikx}+d_{q_{\alpha}}(k)
\bar{\upsilon}_{q}^{\alpha}(k)e^{-ikx}\right).
 \end{eqnarray}
For $\Xi_{cc}^{+}$, $q$=$c, u$.

The contributions from the Pauli-interference(PI) non-spectator
diagrams to the width of $\Xi_{cc}^{+}$ are:
\begin{eqnarray}
\hat{\Gamma}_{PI}^{\Xi_{cc}^{+}}&=&-\frac{2G_{F}^{2}}
{3\pi}\{|V_{ud}|^{2}|V_{cd}|^{2}F_{\mu \nu}(z_{u-},z_{d-}) [N
C_1^2(\mu) \bar{c}\gamma^{\mu}L d \bar{d}{\gamma^\nu}L c
+C_2^2(\mu) \bar{c}^i\gamma^{\mu}L d^j \bar{d}^j{\gamma^\nu}L
c^i\nonumber \\
&&+2C_1(\mu) C_2(\mu) \bar{c}\gamma^{\mu}L d \bar{d}{\gamma^\nu}L
c] +2|V_{cd}|^{2}F_{\mu \nu}(0,z_{l-})\bar{c}\gamma^{\mu}L d
\bar{d}{\gamma^\nu}L c \},
\end{eqnarray}
where $z_{q-}=\frac{m_{q}^{2}}{P_{-}^{2}}(q=u,d,e,\mu)$ and
$P_{-}=p_{c}-p_{d}$. The definition of the function $F_{\mu \nu
}(z_1,z_2)$ is
\begin{eqnarray}
F_{\mu \nu
}(z_1,z_2)&=&-[2(x_2^3-x_1^3)-\frac{3}{2}(2+z_1-z_2)
(x_2^2-x_1^2)+3(x_2-x_1)]P_{-}^2 g_{\mu\nu}\nonumber \\
&&+[2(x_2^3-x_1^3)-3(x_2^2-x_1^2)]P_{-\mu}P_{-\nu},
\end{eqnarray}
where the definitions of $z_1$ and $z_2$ are the same as before.\\
(ii) The inclusive decays of $\Xi_{cc}^{++}$:\\
The non-spectator contribution to the width of $\Xi_{cc}^{++}$come
from the diagrams shown in Fig.2. That is caused by an
interference of the produced $u-$quark from decay of one of the
charm quarks with the $u-$quark in $\Xi_{cc}^{++}$. Here we also
include the $CKM$ suppressed Feynman diagrams. The contribution is
\begin{eqnarray}
\hat{\Gamma}_{PI}^{\Xi_{cc}^{++}}&=&-\frac{2G_{F}^{2}}{3\pi}\{|V_{cs}|^{2}|V_{ud}|^{2}
F_{\mu \nu}(z_{s-},z_{d-})+|V_{cs}|^{2}|V_{us}|^{2}F_{\mu
\nu}(z_{s-},z_{s-})\nonumber \\
&&+|V_{cd}|^{2}|V_{ud}|^{2}F_{\mu \nu}(z_{d-},z_{d-})\}\nonumber \\
&&\{C_{1}^{2}(\mu)\bar{c}^{i}\gamma^{\mu}Lu^{j}\bar{u}^{j}\gamma^{\nu}Lc^{i}+
N C_{2}^{2}(\mu)\bar{u}\gamma^{\mu}L c \bar{c}{\gamma^{\nu}}L u
  +2C_{1}(\mu)C_{2}(\mu)\bar{u}\gamma^{\mu}Lc
\bar{c}L^{\nu} u\},\nonumber \\
\end{eqnarray}
where $z_{-}=\frac{m_{q}^{2}}{P_{-}^{2}}(q=s,d)$, $P_{-}=p_{c}-p_{u}$.\\
(iii) For the inclusive decays of $\Omega_{cc}^{+}$:\\
The non-spectator contributions for $\Omega_{cc}^{+}$ not only
come from the Pauli interference of the $s-$ quark produced in the
non-leptonic, but also from the semi-leptonic decay of the charm
quarks with the $s-$quark in $\Omega_{cc}^{+}$, the later one is
suggested by Voloshin et al.\cite{Voloshin}. As above, here we
include the CKM suppressed WE non-spectator diagrams. The WE
non-spectator contribution to the width $\Omega_{cc}^{+}$ is
\begin{eqnarray}
\hat{\Gamma}_{WE}^{\Omega_{cc}^{+}}&=&\frac{2G_{F}^{2}}{\pi}
|V_{us}|^{2} |V_{cs}|^{2}C(z_{u+},z_{s+}) P_{+}^{2}\nonumber \\
&&\{[C_{1}^{2}(\mu)+C_{2}^{2}(\mu)] \bar{c}\gamma^{\mu}L c
\bar{s}{\gamma_\mu}L s+ 2 C_{1}(\mu)C_{2}(\mu)
\bar{c}\gamma^{\mu}L s \bar{s}{\gamma_\mu}L c\}, \nonumber \\
\end{eqnarray}
where $z_{q+}=\frac{m_{q}^{2}}{P_{+}^{2}}$, $q=u,s$ and
$P_{+}=p_{c}+p_{s}$. \\
The PI non-spectator contribution to the width of
$\Omega_{cc}^{+}$ is
\begin{eqnarray}
\hat{\Gamma}_{PI}^{\Omega_{cc}^{+}}&=&-\frac{2G_{F}^{2}}{3\pi}
\{|V_{cs}|^{2} |V_{ud}|^{2}F_{\mu \nu}(z_{u-},z_{d-})
+|V_{cs}|^{2} |V_{us}|^{2}F_{\mu \nu}(z_{u-},z_{s-})\} \nonumber \\
&&\{NC_{1}^{2}(\mu)\bar{c}\gamma^{\mu}L s \bar{s}\gamma^{\nu}L
c+C_{2}^{2}(\mu) \bar{c}^{i}\gamma^{\mu}Ls^{j}
\bar{s}^{j}{\gamma^\nu}L c^{i}
+2C_{1}(\mu)C_{2}(\mu)\bar{c}\gamma^{\mu}L s \bar{s}\gamma^{\nu}L c\} \nonumber \\
&&-2\frac{2G_{F}^{2}}{3\pi}|V_{cs}|^{2}F_{\mu\nu}(0,z_{l-})\bar{c}\gamma^{\mu}L
s\bar{s}\gamma^{\nu}L c,
\end{eqnarray}
where $z_{q-}=\frac{m_{q}^{2}}{P_{-}^{2}}$, $q=u,d,s,e,\mu$ and
$P_{-}=p_{c}-p_{s}$. Sandwiching the operators between initial and
final $\Xi_{cc}^{+}$, $\Xi_{cc}^{++}$, $\Omega_{cc}^{+}$ states,
we obtain the hadronic matrix elements:
\begin{eqnarray}
&&\Gamma_{WE/PI}^{\Xi_{cc}^{+}}=\langle\Xi_{cc}^{+}(\mathbf{P}=0,s)
|\hat{\Gamma}_{WE/PI}^{\Xi_{cc}^{+}}|\Xi_{cc}^{+}(\mathbf{P}=0,s)\rangle
\nonumber \\
&&\Gamma_{PI}^{\Xi_{cc}^{++}}=\langle\Xi_{cc}^{++}(\mathbf{P}=0,s)
|\hat{\Gamma}_{PI}^{\Xi_{cc}^{++}}|\Xi_{cc}^{++}(\mathbf{P}=0,s)\rangle
\nonumber \\
&&\Gamma_{WE/PI}^{\Omega_{cc}^{+}}=\langle\Omega_{cc}^{+}(\mathbf{P}=0,s)
|\hat{\Gamma}_{WE/PI}^{\Omega_{cc}^{+}}|
\Omega_{cc}^{+}(\mathbf{P}=0,s)\rangle.
\end{eqnarray}
\section{The hadronic matrix elements}
Because the hadronic matrix elements are fully determined by the
non-perturbative QCD effects which cannot be reliably evaluated at
present yet, we need to invoke concrete phenomenological models to
carry out the computations. In this work, we adopt a simple
non-relativistic model, i.e. the harmonic oscillator\cite{Le
Yaouanc}. This model has been widely employed in similar
researches\cite{M.Oda,A.Hosaka,Bonnaz,Barnes,H.Y.Cheng2,Amundson}.
In fact, an advantage of the calculations of the lifetimes of heavy
hadrons is that one does not need to deal with the hadronization
process of lighter products (quarks or even gluons) and the heavy
hadrons can be well described by such simple non-relativistic
models, and the results are relatively reliable
than for light hadron decays.\\

(i)The inclusive decays of $\Xi_{cc}^{+}$:\\
In the harmonic oscillator model, the wavefunction of $\Xi^+_{cc}$
is expressed as $|\Xi^+_{cc}\rangle$ and
\begin{eqnarray}
|\Xi_{cc}^{+}(\mathbf{P},s)\rangle&=&A_{B}\sum_{color,spin}\chi_{spin,flavor}\varphi_{color}\nonumber \\
&&\int
d^{3}p_{\rho}d^{3}p_{\lambda}\Psi_{\Xi_{cc}^{+}}(\mathbf{p_{\rho},p_{\lambda}})
|c_{i}(\mathbf{p}_{q_{1}},s_{q_{1}}),c_{j}(\mathbf{p}_{q_{2}},s_{q_{2}}),d_{k}(\mathbf{p}_{q_{3}},s_{q_{3}})\rangle.
\nonumber \\
\end{eqnarray}
The normalization condition for
$|\Xi_{cc}^{+}(\mathbf{P},s)\rangle$ is
\begin{eqnarray}
&&\langle\Xi_{cc}^{+}(\mathbf{P},s)|\Xi_{cc}^{+}(\mathbf{P'},s')\rangle=(2\pi)^{3}\frac{M_{\Xi_{cc}}}
{\omega_{P}}\delta^{3}(\mathbf{P}-\mathbf{P'})\delta_{s,s'},
\end{eqnarray}
where $\chi_{spin,flavor}$£¬$\varphi_{color}$  are the spin-flavor
and color wavefunctions respectively. Their explicit expressions
are
\begin{eqnarray}
&&\chi_{s=\frac{1}{2},flavor}=\frac
{1}{\sqrt{6}}(2|c_{\uparrow}c_{\uparrow}d_{\downarrow}\rangle-|c_{\uparrow}c_{\downarrow}d_{\uparrow}\rangle-|c_{\downarrow}c_{\uparrow}d_{\uparrow}\rangle)\\
&&\varphi_{color}=\frac{1}{\sqrt{6}}\epsilon_{ijk}.
\end{eqnarray}
$A_{B}$ is the normalization constant. The spatial wavefunction
$\Psi_{\Xi_{cc}^{+}}$ is a three-body harmonic oscillator
wavefunction and expressed as
\begin{eqnarray}
&&\Psi_{\Xi_{cc}^{+}}=\mathrm{exp}(-\frac{\mathbf{p_{\rho}^{2}}}{2a_{\rho}^{2}}-\frac
{\mathbf{p}_{\lambda}^{2}}{2a_{\lambda}^{2}}).
\end{eqnarray}
Here $a_{\rho}$ and $a_{\lambda}$ parameters reflecting the
non-perturbative effects. In the above expressions, the Jacobi
transformations of $\mathbf{p_{1}}$, $\mathbf{p_{2}}$,
$\mathbf{p_{3}}$ which are the momenta of the three valence quarks
$ccd$, and variables $\mathbf{p_{\rho}}$, $\mathbf{p_{\lambda}}$,
$\mathbf{P}$ are
\begin{eqnarray}
\mathbf{p_{\rho}}=\frac {\mathbf{p_{1}-p_{2}}}{\sqrt{2}},
\mathbf{p_{\lambda}}=\frac {\mathbf{p_{1}+p_{2}}-\frac
{2m_{c}}{m_{d}}\mathbf{p_{3}}}{\sqrt{2 \frac
{2m_{c}+m_{d}}{m_{d}}}}, \mathbf{P}=\mathbf{p_{1}+p_{2}+p_{3}}.
\end{eqnarray}

We choose the center-of-mass frame of $\Xi^+_{cc}$, i.e.
($\mathbf{P}$=0) to calculate the hadronic matrix elements.
Substituting the four-quark operators into the expressions, we
obtain the non-spectator WE contributions to the width of
$\Xi^+_{cc}$ as
\begin{eqnarray}
\Gamma_{WE}^{\Xi_{cc}^{+}} &=& 64\pi^2 G_F^2 P_+^2(|V_{cs}|^{2}
|V_{ud}|^{2}C(z_{s+},z_{u+})
+|V_{cd}|^{2} |V_{ud}|^{2}C(z_{u+},z_{d+}))(C_1(\mu)-C_2(\mu))^2\nonumber \\
&&|A_B|^2[2(1+{2m_c\over m_d})]^{3/2} \sum_{spin}\int
d^3\mathbf{p}_\rho d^3\mathbf{p}_\lambda
d^3\mathbf{p'_\rho}\nonumber \\
&&\mathrm{exp}[-{\mathbf{p}_\rho^2\over
2a^2_\rho}-{\mathbf{p}_\lambda^2\over
2a^2_\lambda}]\mathrm{exp}[-{\mathbf{p}_\rho^2\over
2a^2_\rho}-{(\mathbf{p}_\lambda+\sqrt{1+{2m_c\over
m_d}}(\mathbf{p}_\rho-\mathbf{p'_\rho}))^2\over 2a^2_\lambda
}]\bar{u}_{c}\gamma_\mu L u_{c} \bar{u}_{d}\gamma^\mu L
u_{d},\nonumber \\
\end{eqnarray}
and the PI contribution is
\begin{eqnarray}
\Gamma_{PI}^{\Xi_{cc}^{+}} &=& -{64\over 3}\pi^2 G_F^2
\{|V_{cd}|^{2} |V_{ud}|^{2}
F_{\mu\nu}(z_{u-},z_{d-})[-NC_1^2(\mu)+C_2^2(\mu)-2C_1(\mu)C_2(\mu)]\nonumber \\
&&-2|V_{cd}|^{2}F_{\mu\nu}(0,z_{l-})\} |A_B|^2[2(1+{2m_c\over
m_d})]^{3/2} \sum_{spin}\int d^3\mathbf{p}_\rho
d^3\mathbf{p}_\lambda
d^3\mathbf{p'_\rho}\nonumber \\
&&\mathrm{exp}[-{\mathbf{p}_\rho^2\over
2a^2_\rho}-{\mathbf{p}_\lambda^2\over
2a^2_\lambda}]\mathrm{exp}[-{\mathbf{p}_\rho^2\over
2a^2_\rho}-{(\mathbf{p}_\lambda+\sqrt{1+{2m_c\over
m_d}}(\mathbf{p}_\rho-\mathbf{p'_\rho}))^2\over 2a^2_\lambda
}]\bar{u}_{c}\gamma^\mu L u_{d}\ \bar{u}_{d}\gamma^\nu L
u_{c},\nonumber \\
\end{eqnarray}
where the sum over spin means a sum over the polarizations of the
three valence quarks of $\Xi^+_{cc}$ with their corresponding C-G
coefficients in the spin-flavor wavefunction. $u_{q}$,
$\bar{u}_{q}$ denote the Dirac spinors of free quarks $q$ and the
expression is
\begin{eqnarray}
u_{q}=\sqrt{\frac{E_{q}+m_{q}}{2m_{q}}}
\left(%
\begin{array}{c}
  1 \\
  \frac {\mathbf{\sigma} \cdot{\mathbf{p}}}{E_{q}+m_{q}} \\
\end{array}%
\right)\chi\\
\bar{u}_{q}=\sqrt{\frac{E_{q}+m_{q}}{2m_{q}}}\chi^{\dagger}
\left(%
\begin{array}{cc}
  1 & - \frac {\mathbf{\sigma} \cdot{\mathbf{p}}}{E_{q}+m_{q}}\\
\end{array}%
\right)
\end{eqnarray}
in our case $q$ denotes $c$ and $d$ quarks.\\
(ii)The inclusive decays of $\Xi_{cc}^{++}$:\\
The contribution from the PI non-spectator diagrams to
 the width of $\Xi_{cc}^{++}$ is
\begin{eqnarray}
\Gamma_{PI}^{\Xi_{cc}^{++}}&= &-{64\over 3}\pi^2 G_F^2
\{|V_{cs}|^{2} |V_{ud}|^{2} F_{\mu\nu}(z_{s-},z_{d-})+
|V_{cs}|^{2} |V_{us}|^{2} F_{\mu\nu}(z_{s-},z_{s-})\nonumber \\
&&+|V_{cd}|^{2} |V_{ud}|^{2} F_{\mu\nu}(z_{d-},z_{d-})\}
(C_1^2(\mu)-N C_2^2(\mu)-2C_1(\mu)C_2(\mu))|A_B|^2[2(1+{2m_c\over
m_u})]^{3/2} \nonumber \\
&&\sum_{spin}\int d^3\mathbf{p}_\rho d^3\mathbf{p}_\lambda
d^3\mathbf{p'_\rho}\mathrm{exp}[-{\mathbf{p}_\rho^2\over
2a^2_\rho}-{\mathbf{p}_\lambda^2\over
2a^2_\lambda}]\mathrm{exp}[-{\mathbf{p}_\rho^2\over
2a^2_\rho}-{(\mathbf{p}_\lambda+\sqrt{1+{2m_c\over
m_u}}(\mathbf{p}_\rho-\mathbf{p'_\rho}))^2\over 2a^2_\lambda
}]\nonumber \\
&&\bar{u}_{c}\gamma^\mu L u_{u}\ \bar{u}_{u}\gamma^\nu L u_{c}.
\end{eqnarray}
Similar to the case of $\Xi^{+}_{cc}$, the sum over spin means a
sum of the polarizations of the three valence quarks of
$\Xi^{++}_{cc}$ with their C-G coefficients. One only needs to
replace $u$ by $d$ in $\mathbf{p_{\rho}}$, $\mathbf{p_{\lambda}}$
and other expressions are similar to that for $\Xi^{+}_{cc}$.\\
(iii)The inclusive decays of $\Omega_{cc}^{+}$:\\
The contribution from the W-boson exchange(WE) non-spectator
diagrams to the width of $\Omega_{cc}^{+}$ is
\begin{eqnarray}
\Gamma_{WE}^{\Omega_{cc}^{+}} &=& 64\pi^2 G_F^2 P_+^2 |V_{us}|^{2}
|V_{cs}|^{2}C(z_{u+},z_{s+})
  (C_1(\mu)-C_2(\mu))^2\nonumber \\
&&|A_B|^2[2(1+{2m_c\over m_s})]^{3/2} \sum_{spin}\int
d^3\mathbf{p}_\rho d^3\mathbf{p}_\lambda
d^3\mathbf{p'_\rho}\nonumber \\
&&\mathrm{exp}[-{\mathbf{p}_\rho^2\over
2a^2_\rho}-{\mathbf{p}_\lambda^2\over
2a^2_\lambda}]\mathrm{exp}[-{\mathbf{p}_\rho^2\over
2a^2_\rho}-{(\mathbf{p}_\lambda+\sqrt{1+{2m_c\over
m_s}}(\mathbf{p}_\rho-\mathbf{p'_\rho}))^2\over 2a^2_\lambda
}]\bar{u}_{c}\gamma_\mu L u_{c} \bar{u}_{s}\gamma^\mu L
u_{s},\nonumber \\
\end{eqnarray}
whereas that from the Pauli-interference(PI) non-spectator
diagrams is
\begin{eqnarray}
\Gamma_{PI}^{\Omega_{cc}^{+}} &=& -{64\over 3}\pi^2 G_F^2
\{[|V_{cs}|^{2}|V_{ud}|^{2}F_{\mu\nu}(z_{u-},z_{d-})+|V_{cs}|^{2}|V_{us}|^{2}F_{\mu\nu}(z_{u-},z_{s-})]\nonumber \\
&&[-N C_1^2(\mu)+ C_2^2(\mu)-2C_1(\mu)C_2(\mu)]-
 2|V_{cs}|^{2}F_{\mu\nu}(0,z_{l-})\}|A_B|^2[2(1+{2m_c\over m_s})]^{3/2}\nonumber \\
&&\sum_{spin}\int d^3\mathbf{p}_\rho d^3\mathbf{p}_\lambda
d^3\mathbf{p'_\rho}\mathrm{exp}[-{\mathbf{p}_\rho^2\over
2a^2_\rho}-{\mathbf{p}_\lambda^2\over
2a^2_\lambda}]\mathrm{exp}[-{\mathbf{p}_\rho^2\over
2a^2_\rho}-{(\mathbf{p}_\lambda+\sqrt{1+{2m_c\over
m_s}}(\mathbf{p}_\rho-\mathbf{p'_\rho}))^2\over 2a^2_\lambda
}]\nonumber \\
&& \bar{u}_{c}\gamma^\mu L u_{s} \bar{u}_{s}\gamma^\nu L u_{c}.
\end{eqnarray}
The sum over polarizations is similar to that for $\Xi^{+}_{cc}$ and $\Xi^{++}_{cc}$. \\
\section{Input parameters and Numerical results}
To obtain the decay amplitudes, we adopt the input parameters as
follows\cite{Kiselev,pdg}: $G_{F}=1.166 \times10^{-5}
\mathrm{GeV}^{-2}$, $|V_{cs}| = 0.9737$, $|V_{ud}| = 0.9745$,
$C_1(m_c)=1.3$, $C_2(m_c)=-0.57$, $m_c = 1.60$ GeV, $m_s = 0.45$
GeV, $m_u = m_d = 0.3$ GeV, $m_s^{*} = 0.2$GeV, $m_u^{*} = m_d^{*}
= 0$, $M_{\Xi^+_{cc}} = M_{\Xi^{++}_{cc}} = 3.519$ GeV,
$M_{\Omega^+_{cc}} = 3.578$ GeV,
$M_{\Xi^{+*}_{cc}}-M_{\Xi^+_{cc}}=M_{\Xi^{++*}_{cc}}-M_{\Xi^{++}_{cc}}
=M_{\Omega^{+*}_{cc}}-M_{\Omega^+_{cc}}=0.132$ GeV. Here
$m_{q^{*}}$ denotes the current quark mass of flavor $q$.

The non-perturbative parameters $a_{\rho}$, $a_{\lambda}$ in the
harmonic oscillator wavefunctions are selected as follows: for
$J/\psi$, in ref.\cite{Le Yaouanc}, $a_{\rho}^{2}= 0.33
\mathrm{GeV}^{2}$,  for $D-$mesons, $a_{\rho}^{2}=
0.25\mathrm{GeV}^{2}$. For the doubly charmed baryons, because
$a_{\rho}$ reflects the coupling between two charm quarks, we set
it to be the same as that for $J/\psi$. $a_{\lambda}$ reflects the
coupling of the light quark with these two charm quark, thus we
can reasonably set it to be the same as $a_{\rho}$ in D-mesons.

With these parameters as input, the lifetimes of the doubly charmed
baryons can be evaluated out (see TABLE.I), if the non-spectator
effects are taken into account. \\
\begin{table}
\caption{The numerical results about the contributions from the
different components and the evaluated lifetime for the doubly
charmed baryons. For a comparison, in the following table, we list
the corresponding lifetimes predicted by the authors of
ref.\cite{Kiselev} where the diquark picture was employed. It is
noted that in ref.\cite{Kiselev}, the authors used various input
parameters and obtained slightly diverse results, we take average
values of the numbers in the table. There is only one datum for
the lifetimes on $\tau_{\Xi_{cc}^+}$ given by the SELEX
collaboration which is also listed the table.} \vspace{2mm}
\begin{tabular}{|c|c|c|c|c|c|c|}
  \hline
  $\Xi^+_{cc}$  & $\Gamma_{spec}(10^{-12}\mathrm{GeV})$ & $\Gamma^{WE}_{non}
  ( 10^{-13}\mathrm{GeV})$  & $\Gamma^{PI}_{non}( 10^{-15}
  \mathrm{GeV})$  & $\tau_{\Xi_{cc}^+}(\mathrm{ps})$  & $\tau_{\Xi_{cc}^+}(\mathrm{ps})$ in ref.\cite{Kiselev} & exp($\mathrm{ps}$)\\
  \hline
  &2.01 & 6.43 & -3.36 & 0.25  & 0.19 & 0.033\\
 \hline
  $\Xi^{++}_{cc}$& $\Gamma_{spec}( 10^{-12}\mathrm{GeV})$& & $\Gamma^{PI}_{non}
  ( 10^{-12}\mathrm{GeV})$  & $\tau_{\Xi_{cc}^{++}}(\mathrm{ps})$ & $\tau_{\Xi_{cc}^{++}}(\mathrm{ps})$ in ref.\cite{Kiselev} &\\
  \hline
&  2.01 & & -1.02 & 0.67 & 0.52 & $-$\\
 \hline
    $\Omega^{+}_{cc}$ & $\Gamma_{spec}( 10^{-12}\mathrm{GeV})$ & $\Gamma^{WE}_{non}( 10^{-14}\mathrm{GeV})$   &$\Gamma^{PI}_{non}
  ( 10^{-12}\mathrm{GeV})$
  & $\tau_{\Omega_{cc}^+}(\mathrm{ps})$ & $\tau_{\Omega_{cc}^{+}}(\mathrm{ps})$ in ref.\cite{Kiselev} & \\
  \hline
& 2.01 &  4.25 & 1.10   & 0.21 & 0.22 & $-$\\
\hline
\end{tabular}
\end{table}

\section{Conclusion and Discussion}

In this work, we evaluate the lifetimes of doubly charmed baryons
with the non-spectator effects being properly taken into account.
As argued in the introduction, to evaluate the lifetimes (the
total widths), only the inclusive processes are concerned, and
then the non-perturbative effects are all from the wavefunctions
of the doubly charmed baryons. Due to existence of the two heavy
charm quarks, the non-relativistic harmonic oscillator model
should apply in this case. Mainly, we carefully calculate the
contribution of non-perturbative effects to the lifetimes in the
model, which are closely related to the bound states of the
baryons.

Our numerical results indicate that the non-spectator
contributions to the lifetimes of $\Xi^+_{cc}$, $\Xi^{++}_{cc}$
and $\Omega^+_{cc}$ are substantial. The non-spectator
contributions to the width of $\Xi^+_{cc}$ are mainly from the WE
diagrams (the PI diagrams which contribute are CKM suppressed),
since the WE contribution is constructive, therefore the lifetime
of $\Xi^+_{cc}$ is much suppressed. By contraries, for
$\Xi^{++}_{c}$ and $\Omega^+_{cc}$, the non-spectator
contributions are mainly from the PI diagrams and the net effect
is destructive. It is noted that for $\Omega^+_{cc}$ there still
are Cabibbo-suppressed WE diagrams, but for $\Xi^{++}_{cc}$ there
are only PI diagrams. Therefore the predicted lifetime of
$\Xi^{++}_{cc}$ is larger than that of other two baryons. We also
employ other values for parameters $a_\rho, a_\lambda$ and find
that the resultant values can vary within 20\% uncertainty.

Our results are
$$ \tau(\Xi^+_{cc})=0.25\; {\rm ps}£¬\;\;\; \tau(\Xi^{++}_{cc})=
0.67\;{\rm ps}£¬\;\;\;\;{\rm and}\;\;\;\; \tau(\Omega^+_{cc})=
0.21\;{\rm ps}.$$

These are generally consistent with the results obtained by
Kiselev et al.\cite{Kiselev} and Guberina et al.\cite{Guberina},
even though they used different models for calculating the
hadronic matrix elements. Concretely, they used the diquark
picture and attributed the non-perturbative effects into the
wavefunction of the diaquark at origin. Kiselev et al. gave
$\tau(\Xi^+_{cc})\sim 0.16-0.22$ ps $\tau(\Xi^{++}_{cc})\sim
0.40-0.65$ ps and $\tau(\Omega^+_{cc})\sim 0.24-0.28$.

Although all the theoretical predictions based on different models
agree with each other, they are obviously one order larger than
the upper limit of the measured value on the lifetime of
$\Xi^+_{cc}$ (0.033 ps) by the SELEX collaboration\cite{SELEX}.
This deviation, as suggested by some authors, may come from
experiments\cite{Kiselev2}. So far the difference between
theoretical predictions and experimental data may imply some
unknown physics mechanisms which drastically change the value, if
the future experiment, say at LHCb, confirms the measurement of
the SELEX. Recently, several groups have studied the possibility
of doubly heavy baryon production at hadron collider LHC and
future linear collider ILC\cite{hadroncollider,linearcollider} and
the effective field theories for two heavy quarks system are also
further investigated\cite{EFT}. We are expecting the new data from
more accurate experiments at LHC and ILC to improve our
theoretical framework and determine if there are contributions
from new physics beyond the standard model.\\

\noindent Acknowledgement: This work is supported by the National
Natural Science Foundation of China.\\

\pagebreak
\begin{figure}[htb]
\begin{center}
\includegraphics[width=15cm]{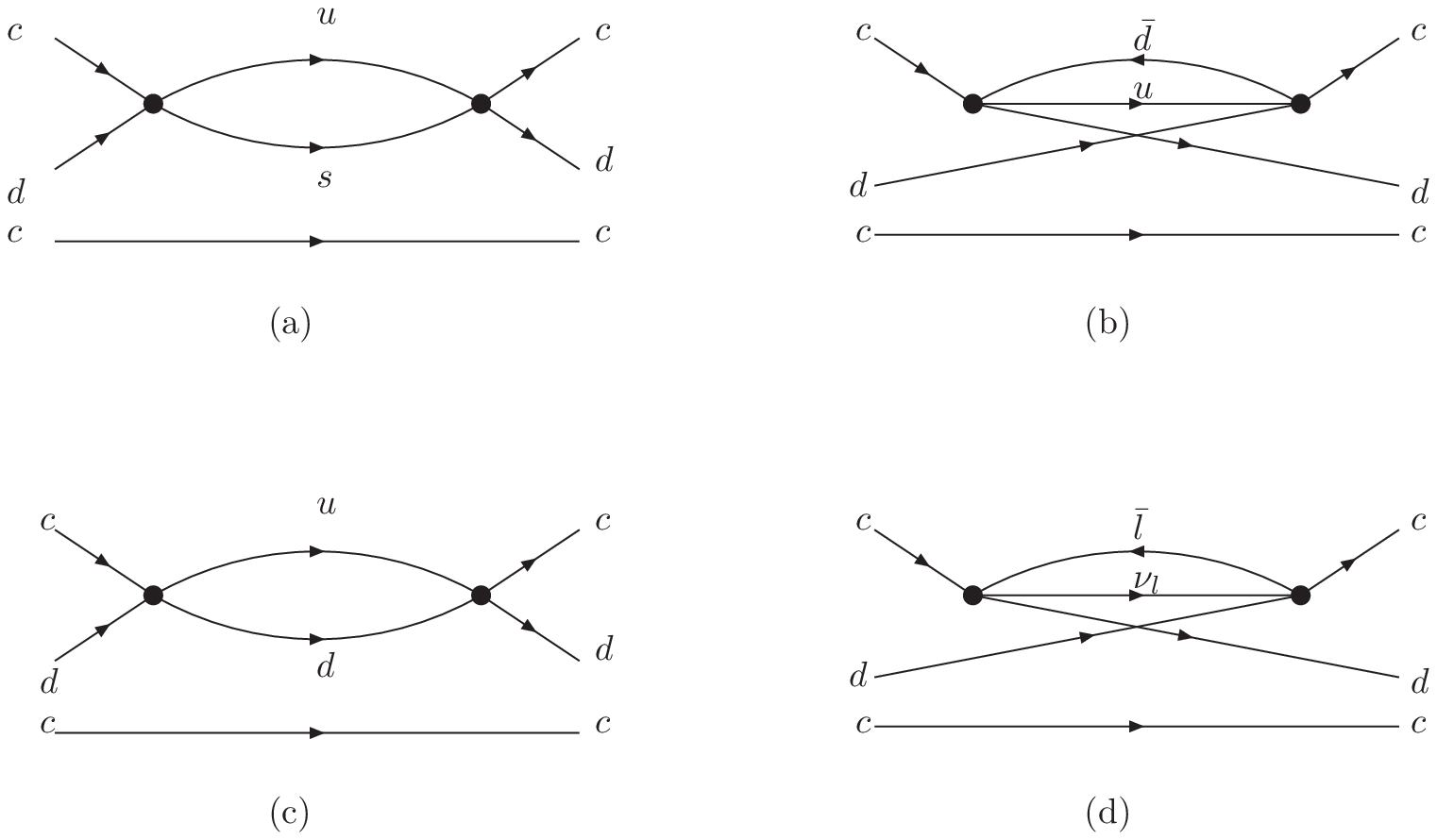}
\label{figure} \caption{non-spectator effects contribution to
lifetime of $\Xi_{ccd}$}
\end{center}
\end{figure}
\pagebreak
\begin{figure}[htb]
\begin{center}
\includegraphics[width=15cm]{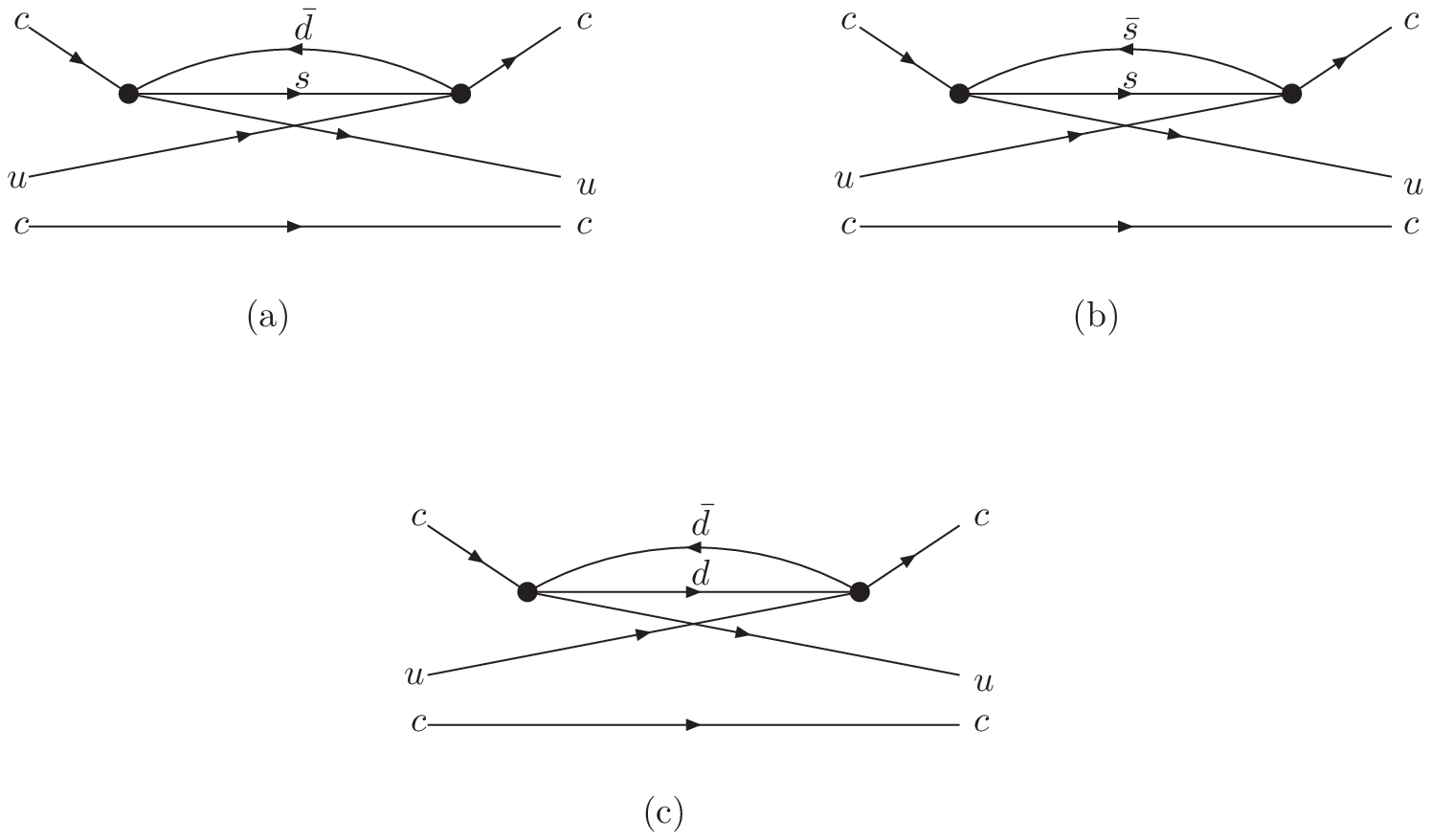}
\caption{non-spectator effects contribution to lifetime of
$\Xi_{ccu}$}
\end{center}
\end{figure}
\pagebreak
\begin{figure}[htb]
\begin{center}
\includegraphics[width=15cm]{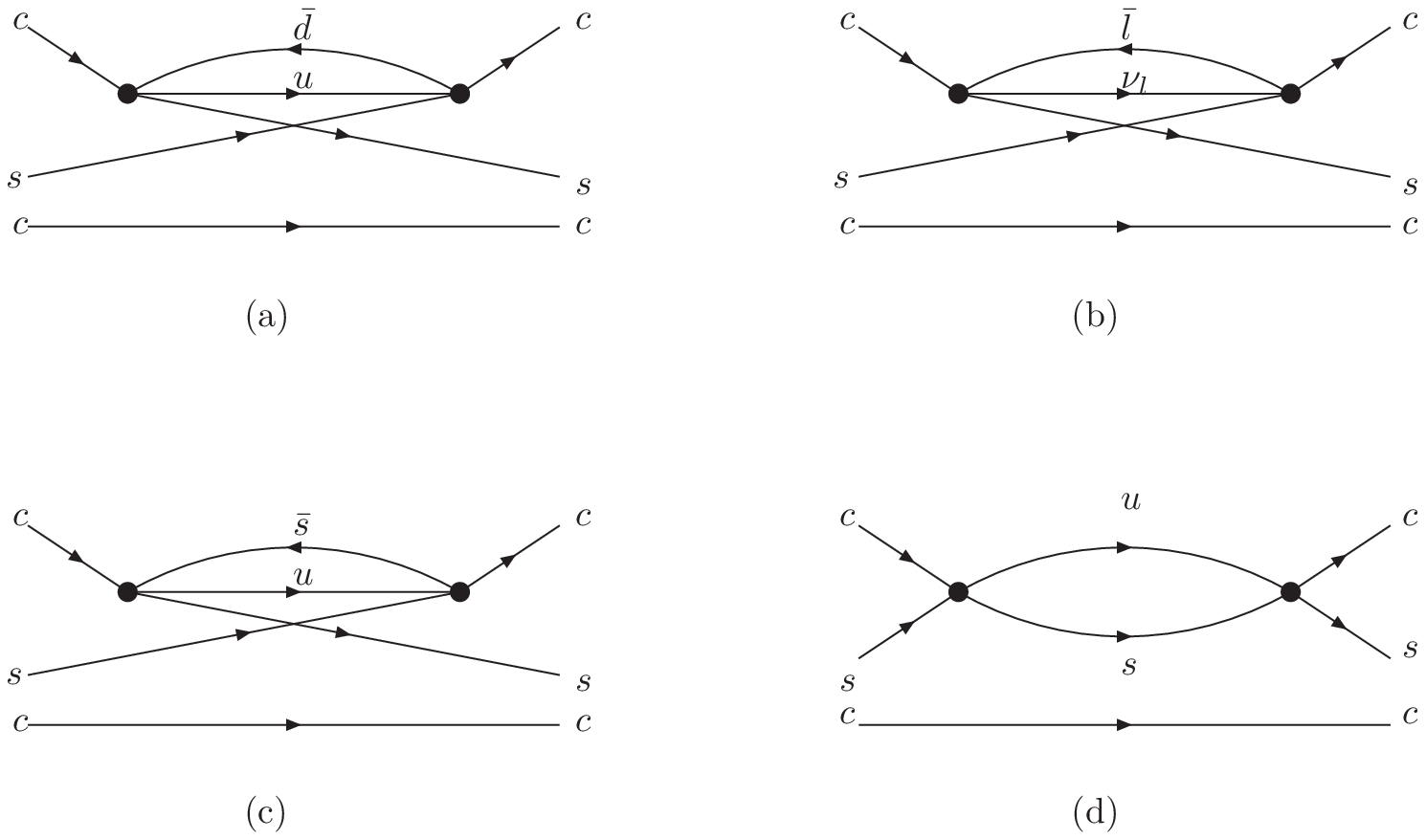}
\caption{non-spectator effects contribution to lifetime of
$\Omega_{ccs}$}
\end{center}
\end{figure}
\end{document}